\documentclass[%
 reprint,
superscriptaddress,
 amsmath,amssymb,
 aps,prd,
floatfix,
10pt
]{revtex4-1}

\usepackage{graphicx}
\usepackage{dcolumn}
\usepackage{bm}
\usepackage{hyperref}
\usepackage[caption=false]{subfig}

\usepackage[dvipsnames]{xcolor}

\DeclareMathOperator{\sech}{sech}

\newcommand{\diff}[1]{\,\mathrm{d}#1}						
\newcommand{\dtot}[2]{\frac{\mathrm{d}#1}{\mathrm{d}#2}}			
\renewcommand{\vec}[1]{\mathbf #1}									

\let\oldsqrt\sqrt
\def\sqrt{\mathpalette\DHLhksqrt}
\def\DHLhksqrt#1#2{%
\setbox0=\hbox{$#1\oldsqrt{#2\,}$}\dimen0=\ht0
\advance\dimen0-0.2\ht0
\setbox2=\hbox{\vrule height\ht0 depth -\dimen0}%
{\box0\lower0.4pt\box2}}

\begin{document}

\preprint{APS/123-QED}

\title{Massive scalar wave packet emission by a charged Black Hole and Cosmic Censorship Conjecture violation}
\author{Rodrigo L. Fernandez}
\affiliation{Universidade Federal do Rio de Janeiro, Instituto de F\'isica, CEP 21941-972, Rio de Janeiro, RJ, Brasil}
\author{Ribamar R. R. Reis}
\affiliation{Universidade Federal do Rio de Janeiro, Instituto de F\'isica, CEP 21941-972, Rio de Janeiro, RJ, Brasil}
\affiliation{Universidade Federal do Rio de Janeiro, Observat\'orio do Valongo, 
CEP 20080-090, Rio de Janeiro, RJ, Brazil}
\author{Sergio E. Jor\'as}
\affiliation{Universidade Federal do Rio de Janeiro, Instituto de F\'isica, CEP 21941-972, Rio de Janeiro, RJ, Brasil}

\begin{abstract}
We study the tunneling probability of a massive ($m_w$) uncharged scalar packet out from a near-extremal, static charged black hole (with mass $M$ and charge $Q\lesssim M$). We show that there is indeed a \textit{net} probability that a massive uncharged particle tunnels out from the black hole so that the final state (with new mass $M'\equiv M-m_w < Q$) does violate the cosmic censorship conjecture. Nevertheless, the typical time for such a black hole to discharge (i.e, to absorb charge $-Q$ from its surroundings and then become neutral) is much smaller than the tunneling time; therefore, the violation is never attained in practice. Even for a completely isolated black hole (should it exist), the standard time dilation near the horizon stretches the typical violation time scale to unobservable values.
\end{abstract}

\keywords{Cosmic censorship conjecture, toy model, semiclassical approach}

\maketitle

\section{Introduction}
The Cosmic Censorship Conjecture (CCC) states that every singularity (except the cosmological one) must appear ``dressed'' in the universe. This statement was introduced by Roger Penrose~\cite{Penrose2002}, meaning that every singularity (except the Big Bang) in the universe must be hidden inside an Event Horizon. Mathematically, this is described by the inequality $M^2 \geqslant Q^2 + a^2$ (in geometrized unit system), with $M$ being the mass of the Black Hole (BH), $Q$ its charge and $a\equiv J/M$ its specific angular momentum. Essentially, these three quantities determine uniquely a BH, as stated by the \textit{no-hair theorem}~\cite{MTW}.

The quest for a definite proof of the CCC remains for both physicists and mathematicians. For physicists, it imposes the existence of a ``censor'' that forbids any singularity to become visible for any outside observer in the universe, hiding it behind the Event and Cauchy Horizons (EH and CH, respectively). For mathematicians, on the other hand, it means that the flux of gravitational and/or electromagnetic radiation diverges once they cross the Cauchy Horizon~\cite{1982RSPSA.384..301C,Costa:2019uny}.

Wald~\cite{1974AnPhy..82..548W,Wald:1997wa} (and references therein) and Needham~\cite{PhysRevD.22.791} agreed that there would be no CCC violation classically, whereas Hiscock~\cite{Hiscock1979} claimed that if the CCC was upheld, then strange phenomena would appear. Later on, Hubeny~\cite{PhysRevD.59.064013} obtained a violation of the CCC by classically overcharging a near-extremal Reissner-N\"ordstrom BH and turning it into a naked singularity, while Felice and Yunqiang~\cite{0264-9381-18-7-307} turned a Reissner-N\"ordstrom BH into a naked Kerr-like singularity by overspinning it, to which Hod~\cite{PhysRevD.66.024016} replied saying that not only the effects of backreaction and superradiance must be taken into account, but also the particle's self-energy. But it was within the works of Matsas~\cite{PhysRevLett.99.181301,PhysRevD.79.101502} that the possibility for a particle (actually, a plane wave) to tunnelate the BH's potential barrier via a purely quantum effect and violate the CCC that brought new light to the discussion, a consideration to which Hod~\cite{PhysRevLett.100.121101} strongly disagrees. Since then, the question of tunneling the potential barrier --- i.e., a quantum violation of the CCC --- was carried on by other researches and it has been, for the past few years, the main line of research on this topic~\cite{PhysRevD.84.104021,PhysRevD.84.027501,PhysRevD.88.064043,PhysRevD.96.044043}. 
Other works have investigated the role of barrier tunneling in BHs, e.g, the collapse of a charged shell  \cite{PhysRevD.80.124027} --- which would by itself violate the CCC --- or the quantum nature of the horizon \cite{CASADIO201568} --- the latter with no connection to the CCC.

In the present work we take a step further into the possibility for a particle to be emitted from the BH via quantum tunneling process. We first consider a massive scalar field being absorbed by a static, charged BH. Solving the Klein-Gordon equation for the given scalar field, we find a Schr\"odinger-like equation for the radial part (with an effective potential) that has no analytical solution. To circumvent this problem, instead of using only a numerical approach or low (or high) frequency approximation, we also propose a toy model that is as close as possible to the actual effective potential and yields an analytical solution to the aforementioned equation. In the asymptotic limits, we recover the expected plane waves, that allows us to define the reflection and transmission rates and identify the latter (up to a constant factor) as the absorption probability. We proceed to build a Gaussian wave packet from the incoming plane waves as a semiclassical representation of a particle and calculate the absorption probability for such packet. Using the symmetry of the problem we can further relate the transmission rate with the emission probability (that being allowed due a parity transformation of the reflection and transmission rates) and study the probability for the BH to emit a neutral particle with mass $m_w$. If the emitted particle is such that the BH's new mass $M'\equiv M-m_w < Q$, then we have a violation of the CCC, that is, a naked singularity.

The problem for the absorption of a scalar field by a BH has already been discussed for numerous papers throughout the past years~\cite{PhysRevD.89.104053,PhysRevD.96.044043,PhysRevD.84.104021}, but as for a wave packet is a brand new discussion. Few attempts were proposed in the past 30 years~\cite{vishveshwara1970scattering,hawking1975}, and very little has been done about it. We now proceed to explore this problem by building Gaussian wave packets to represent particles being absorbed/emitted by the BH.

We have also used the Pr\"ufer method for a numerical calculation of the phase-shift, as shown in reference~\cite{Glampedakis:2001cx}, to validate our toy model method. 
In a nutshell, this approach focus on the phase of the wave function and provides more robust numerical results, as we will see below.

Throughout this paper, we will set the metric signature as $-,+,+,+$, and use the geometrized system of units where $G=c=1$. We also represent every quantity here as being normalized to the BH's mass.

\section{Basic Equations}
We will study the absorption of a massive scalar field $\phi(t,\vec{r})$ with mass $m_w=\mu\hbar$ by a Reissner-N\"ordstrom (RN) BH of mass $M$ and charge $Q < M$. 

The Klein-Gordon equation reads
\begin{equation}\label{eq:KG}
    \left(\nabla_\nu\nabla^\nu - \mu^2\right)\Phi(t,\vec{r}) = 0
\end{equation}
where $\nabla_\nu$ is the covariant derivative, and the metric describing an empty spacetime structure outside the BH is given by
\begin{multline}\label{eq:RNlinelem}
\diff{s}^2 = 
    -\left(1 - \frac{2M}{r} + \frac{Q^2}{r^2}\right)\diff{t}^2 
    + \frac{1}{1 - \displaystyle\frac{2M}{r} + \frac{Q^2}{r^2}}\diff{r}^2 \\[.5em]
    + r^2\diff\theta^2
    + r^2\sin^2\theta\diff{\varphi}^2
\end{multline}
Since the metric is spherically symmetric, by means of the separation of variables for the massive scalar field where $\phi(t,\vec{r}) = R(r)\Theta(\theta)e^{i(m\varphi-\omega t)}$, with $m\in\mathbb{Z}$ being the orbital parameter and $\omega$ the field's frequency, we get for the angular sector $\Theta(\theta)$ of the Klein-Gordon eq.~\eqref{eq:KG} the expression bellow:
\begin{equation}\label{eq:KGAngular}
	\dtot{^2\Theta}{\theta^2} + \cot\theta\dtot{\Theta}{\theta} + \left(A - \frac{m^2}{\sin^2\theta}\right)\Theta = 0, \\[1em]
\end{equation}
where $A$ is the separation constant. The solutions for eq.~\eqref{eq:KGAngular} are easily obtained by the Legendre polynomials $P_\ell(\cos \theta)$, leading to $A=\ell(\ell+1)$ with $\ell$ being the angular parameter and $m$ is bounded to $\ell$ via the inequality $|m|\leqslant\ell$. The radial part $R(r)$ of eq.~\eqref{eq:KG} then reads
\begin{multline}\label{eq:KGRadial}
	r^2\Bigg(1-\frac{2M}{r}-\frac{Q^2}{r^2}\Bigg)\dtot{^2R}{r^2} 
	+ r\Bigg(2-\frac{2M}{r}\Bigg)\dtot{R}{r} \\
		+ \Bigg(
			\frac{\omega^2r^2+[(2M-r)r + Q^2]\mu^2}{1-\displaystyle\frac{2M}{r}-\frac{Q^2}{r^2}} - \ell(\ell+1)
		\Bigg)R = 0
\end{multline}
Since in this system of coordinates the radial function is only defined in the interval $r_+ < r < \infty$, where $r_+ \equiv M + \sqrt{M^2-Q^2}$ is the EH radius, we proceed to a change of variables to the \textit{tortoise coordinates} where
\begin{equation}
    \dtot{r}{r^\star} \equiv \left(1 - \frac{2M}{r} + \frac{Q^2}{r^2}\right)^{-1}
\end{equation}
with $r^\star\in(-\infty,\infty)$. The tortoise coordinate as a function of the radial coordinate is
\begin{multline}
    r^\star(r) = r + \frac{r_+^2}{r_+ - r_-}\log\left(\frac{r}{r_+} + 1\right) \\
		- \frac{r_-^2}{r_+ - r_-}\log\left(\frac{r}{r_-} + 1\right) + C
\end{multline}
where $r_+$ was already identified as the EH radius while $r_- \equiv M - \sqrt{M^2-Q^2}$ is the CH radius, and $C$ being the integration constant \footnote{Without any loss of generality, we may as well set it to zero, but for a full discussion on the choice of $C$ for it to match the Newtonian phase-shift in the asymptotic limit, see reference~\cite{Glampedakis:2001cx}}. Using the tortoise coordinate, the radial part given by eq.~\eqref{eq:KGRadial} then becomes
\begin{equation}\label{eq:SLE}
    \left[\dtot{^2}{r^{\star^2}} + \omega^2 - V_\mathrm{eff}(r)\right]u_{\ell m}(r) = 0,
\end{equation}
where $u_{\ell m}(r) \equiv rR_{\ell m}(r)$ and
\begin{multline}\label{eq:Veff}
    V_\mathrm{eff}(r) \equiv \left(1 - \frac{2M}{r} + \frac{Q^2}{r^2}\right) \\[.5em]
    \times\left(\frac{\ell(\ell+1)}{r^2} + \frac{2M}{r^3} - \frac{2Q^2}{r^4} + \mu^2\right)
\end{multline}
is the effective potential. We point out the (standard) mixed-coordinate representation ($r$ and $r^*$) used in the previous equation.

\subsection{Numerical method}
Eq.~\eqref{eq:SLE} is a \textit{Schrodinger-like equation} and, since it is not analytically solvable for the given effective potential, we rely on numerical methods to get solutions for the problem in this subsection (and postpone the toy model for the next one). 

The aforementioned equation can be written as
\begin{equation}
	\frac{d}{dx}\bigg[P(x) \frac{d}{dx}u(x) \bigg] - Q(x) u(x)=0,
\end{equation}
where $x\equiv  \omega r^*$, $P(x)\equiv 1/\tilde\varpi$ and $Q(x)\equiv V_{\rm eff}(x)/\tilde\varpi $, with $\tilde\varpi \equiv \sqrt{1-{\tilde\mu}^2} $, suitable for the Pr\"ufer method, which we now follow. 

First, we write the wave function as $u(x^\star)=\int G(x') dx'$, i.e, $G(x^\star) \equiv  u'/u$. The function $G(x^\star)$, therefore, obeys the following equation:
\begin{equation}
	G'+ G^2 + (1 - V_{\rm eff}) = 0,
\end{equation}
with $G(x^\star\to-\infty) \sim -i$ as its initial condition (we recall that there is only the transmitted ingoing wave in this region).

In the opposite limit, i.e, $x^\star\to +\infty$, there are both the incident and the reflected waves:
\begin{align}
	u(x^\star)
	&= k \exp(-i \tilde \varpi x^\star) + r \exp(+i \tilde \varpi x^\star)\\[.5em]
	&= B \sin(\tilde\varpi x^\star + \xi).
\end{align}

In order to focus on its phase, we define $\tilde G(x^\star)\equiv  \theta(x^\star) - \tilde\varpi x^\star$. Note that, in the limit $x^\star\to +\infty$, one arrives at $\tilde G\to \xi$. Eq.~\eqref{eq:SLE} is now cast as
\begin{equation}
	\tilde G'+ \bigg(\frac{V_{\rm eff}-\tilde \mu^2}{\tilde\varpi} \bigg)\sin^2(\tilde G + \tilde\varpi x^\star)=0.
\end{equation}
The required ``initial'' condition for the equation above is given by matching $\tilde G$ with $G$ at an (arbitrary) intermediate point $x_o^*$:
\begin{equation}
	\tilde G(x_o^*) = -\tilde\varpi x_o^\star + \frac{i}{2}
	\log\bigg(\frac{G(x_o^*)-i}{G(x_o^*)+i} \bigg).
\end{equation}
The value of $\xi$ defines the scattering matrix $S_\ell$ and the phase shift $\delta_\ell$ by
\begin{align}
	S_\ell 		&\equiv  (-1)^{\ell + 1} \frac{r}{k}\equiv  \exp(2 i \delta_\ell), \quad {\rm where} \\
	\delta_\ell &= \xi + \ell \frac{\pi}{2}.
\end{align}

We have written a C code to calculate the phase shift and proceed to obtain the reflection and transmission rates.

In the next subsection, we will arrive at the same quantities by a different method.

\subsection{Toy Model}
\begin{figure}[tb]
	\centering
    \includegraphics[width=0.95\linewidth]{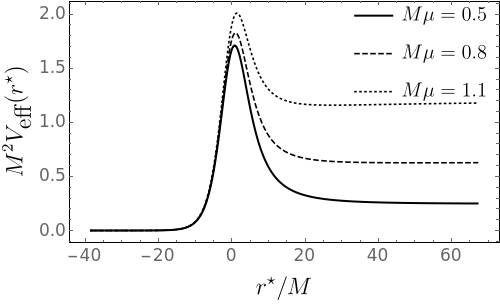}
    \caption{\label{fig:Veff}Plot for the effective potential defined in eq.~\eqref{eq:Veff} for different values for the scalar field's mass $\mu$. The mass $M$ and charge $Q$ of the BH are fixed to be $M=1\,\mbox{m}$ and $Q=0.9M$. The scalar field angular parameter is also fixed at $\ell=5M^2$.}
\end{figure}

As an alternative --- and for supporting the results from the previous approach --- we will propose a \textit{toy model} which consists in modeling the effective potential with functions known to allow analytical solutions to eq.~\eqref{eq:SLE} and that are as close as possible to the actual problem. The effective potential in eq.~\eqref{eq:Veff} is plotted in fig.~\ref{fig:Veff} for different values of the scalar field's mass $\mu$ and fixed values of the scalar field angular parameter $\ell$ and the BH's mass $M$ and charge $Q$.

\begin{figure}[tb]
    \centering
    \includegraphics[width=0.95\linewidth]{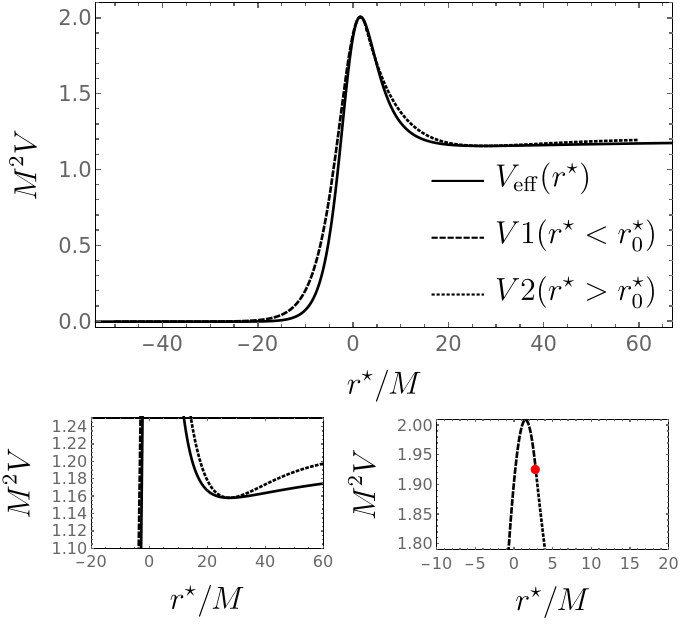}
    \caption{\label{fig:Vtoy}Toy model (dashed and dotted lines) along with the effective potential (solid line) for $M=1\,\mbox{m}$, $Q=0.9M$, $\ell=5M^2$ and $M\mu=1.1$. The local minimum is shown in the left bottom panel, while the junction point $r_0^\star$ is represented by the red dot in the right bottom panel.}
\end{figure}

Our proposal for the toy model will be given by the following asymmetric piecewise function
\begin{multline}
    \!\!\!\!\!\! V_{\rm toy}(r^\star) \equiv
    \begin{cases}
        b_1\sech^2[a_1(r^\star - c_1)], & r^\star \leqslant r_0^\star \\
        (b_2-\mu^2)\{[1-e^{-a_2(r^\star-c_2)}]^2-1\}, & r^\star > r_0^\star
    \end{cases}
\end{multline}
where for $r^\star > r_0^\star$ we have the Morse potential~\cite{PhysRev.34.57} and for $r^\star \leqslant r_0^\star$ we have the P\"oschl-Teller potential~\cite{Skakala2010}. Both functions yield analytical solutions when plugged into eq.~\eqref{eq:SLE}. The parameters $a_i$, $b_i$ and $c_i$, $i=1,2$, are the potential width, height and coordinate position of the extrema (highest and lowest), respectively, all in tortoise coordinates, while $r_0^\star$ is the point where both functions connect.
To set the parameters $a$, $b$ and $c$ we have
\begin{equation}
    c_1 \equiv r^\star(r_\mathrm{max}), \quad c_2 \equiv r^\star(r_\mathrm{min})
\end{equation}
where $r_{\rm max}$ and $r_{\rm min}$ are the values of the $r-$coordinate where the maximum and local minimum of the potential are located, while $b_1$ and $b_2$ given by
\begin{equation}
    b_1 \equiv V(r_{\mathrm{max}}), \quad b_2 \equiv V(r_{\mathrm{min}})
\end{equation}
are the height of the maximum and depth of the local minimum, respectively. For the potential widths $a_1$ and $a_2$, we have chosen $a_1$ to be set by taking the second derivatives of the actual potential and of the toy model with respect to $r^\star$ at each maximum and equating them both, arriving to
\begin{equation}
    a_1 \equiv \left[\sqrt{-\frac{1}{2b_1}\dtot{^2V}{r^{\star^2}}}\right]_{r=r_{\mathrm{max}}}
\end{equation}
while $a_2$ is found together with $r_0^\star$ by imposing the continuity of the potential and its derivative at $r^\star=r_0^\star$,
\begin{align}
    \lim_{r^\star\to r_0^{\star-}} V_{\rm toy}(r^\star) &= \lim_{r^\star\to r_0^{\star+}}V_{\rm toy}(r^\star)
    \label{eq:system1} \\[.5em]
    \lim_{r^\star\to r_0^{\star-}} V'_{\rm toy}(r^\star) &= 
		\lim_{r^\star\to r_0^{\star-}} V'_{\rm toy}(r^\star)
    \label{eq:system2}
\end{align}
where the prime denotes the derivative with respect to $r^\star$. This guarantee that the potential will be smoothly connected at $r_0^\star$. Notice that $c_1 < r_0^\star < c_2$. The toy model is plotted along with the effective potential in fig.~\ref{fig:Vtoy}.

Since the toy model is composed by two analytically integrable solutions to eq.~\eqref{eq:SLE}, it is analytical in every quantity calculated, depending only on the choice of the set of parameters $\{M,Q,\ell,\mu\}$ and their limitations (e.g., the existence of the maximum and/or the local minimum regarding $\mu$, and limits like $Q\to0$ and/or $\ell\to0$). The asymmetry of the toy model was key to best model the effective potential that is asymmetric itself.

The solution to eq.~\eqref{eq:SLE} for a scalar field of frequency $\omega$ and the potential given by the toy model is~\cite{CEVIK20161600,2012arXiv1203.1285P}
\begin{multline}\label{eq:sol1}
    u^{(1)}_{\ell m}(r^\star)
    = \alpha_1\left(\frac{1+y}{1-y}\right)^{\lambda/2}
		{}_2F_1\left(\nu,1-\nu,1+\lambda;\frac{1+y}{2}\right) \\[.7em]
    + \beta_1 \left[\frac{4}{(1+y)(1-y)}\right]^{\lambda/2} \\[.7em]
	\times {}_2F_1\left(\nu-\lambda,1-\nu-\lambda,1-\lambda;\frac{1+y}{2}\right)
\end{multline}
for $r^\star \leqslant r_0^\star$, where
\begin{multline}
    \lambda \equiv  \frac{i\omega}{a_1}, \quad \nu \equiv  \frac{1}{2}\left(1+\sqrt{1-\frac{4b_1}{a_2^2}}\right), \\[.5em]
    y \equiv  \tanh[a_1(r^\star-c_1)]
\end{multline}
and ${}_2F_1(a,b,c;x)$ is the Gaussian Hypergeometric function. For $r^\star > r_0^\star$,
\begin{multline}\label{eq:sol2}
    u^{(2)}_{\ell m}(r^\star) =
    e^{-z/2}\left\{
		\alpha_2 z^\eta {}_1F_1\left(\frac{1}{2}+\eta-\zeta,1+2\eta;z\right) \right.\\[.7em]
		\left. + \beta_2 z^{-\eta} {}_1F_1\left(\frac{1}{2}-\eta-\eta,1-2\eta;z\right)
	\right\}
\end{multline}
with
\begin{equation}
    \eta \equiv  \frac{i\omega}{a_2}, \quad \zeta \equiv  \frac{\sqrt{b_2}}{a_2}, \quad z \equiv  2\zeta e^{-a_2(r^\star-c_2)}
\end{equation}
where ${}_1F_1(a,b;x)$ is the Kummer's function. The coefficients $\alpha_1$, $\alpha_2$, $\beta_1$ and $\beta_2$ are constants to be numerically determined by the conditions that the solutions are equal at the junction point $r_0^\star$, and so are their derivatives. The complete solution to eq.~\eqref{eq:SLE} is then
\begin{equation}
    u_{\ell m}(r^\star) =
    \begin{cases} 
        u^{(1)}_{\ell m}(r^\star), &r^\star \leqslant r_0^\star \\[.5em]
        u^{(2)}_{\ell m}(r^\star), &r^\star > r_0^\star
    \end{cases}
\end{equation}
and it is not hard to see that in the limit $r^\star\to\pm\infty$, given the behavior of the Hypergeometric functions~\cite{Olver:2010:NHM:1830479}, we recover plane waves, as expected:
\begin{equation}\label{eq:asymexp}
    u_{\ell m}(r^\star) \sim
    \begin{cases} 
        ke^{-i\varpi r^\star} + re^{+i\varpi r^\star}, &r^\star \to +\infty \\[.5em]
        te^{-i\omega r^\star}, &r^\star \to -\infty
    \end{cases}    
\end{equation}
where $\varpi \equiv \sqrt{\omega^2-\mu^2}$. In the asymptotic expansion, the coefficients $k$, $r$ and $t$ will depend on the wave's amplitudes (that is, $\alpha_i$ and $\beta_i$), the field's orbital parameter $\ell$, and the BH's mass $M$ and charge $Q$. Notice that the condition of no-outgoing wave from $r^\star\to-\infty$ was already used, i.e., the coefficient of the plane wave $e^{+i\omega r^\star}$ was already set to zero (i.e., there are no waves outgoing from the BH in this case).

The problem is similar to a quantum-tunneling problem, where we have an incoming, a reflected and a transmitted waves (although in the standard tunneling problem there may be an outgoing wave from the left side, given the appropriate boundary conditions), thus we might as well use that piece of information to give meaning to the quantities
\begin{equation}
    R_w \equiv  \left|\frac{r}{k}\right|^2, \quad T_w \equiv  \left|\frac{t}{k}\right|^2
\end{equation}
as the Reflection and Transmission rates, respectively,  for the given scalar field of frequency $\omega$. With the conservation of flux we get the relation between $T_w$ and $R_w$ to be $R_w+|\omega/\varpi|T_w=1$ (the case of an uncharged particle shows no superradiance where the reflection rate could be greater than unity). Also, we may calculate the scattering matrix $S_\ell$ and the phase-shift $\delta_\ell$ as usual:
\begin{equation}
    S_\ell \equiv  (-1)^{\ell+1}\frac{r}{k} \equiv e^{2i\delta_\ell}
\end{equation}

Our toy model is in agreement with the results from the Pr\"ufer method in the frequency range and in the parameter set where both are valid, giving us the confidence to proceed in our calculations.

From now on, we focus on the results from toy model, which can be extended to smaller frequency values.

\section{Wave packet}
We will build the Gaussian wave packet in the position space as the Fourier transform of an also Gaussian wave packet in the frequency space, that is
\begin{align}
    \psi(t,r^\star) &= \int_0^\infty \tilde{\psi}(\omega)e^{-i\omega (t+r^\star)} \diff{\omega}, \\
    \tilde{\psi}(\omega) &= \frac{1}{(2\pi\tilde{\sigma}^2)^{1/4}}\exp\left[-\frac{(\omega-\omega_0)^2}{4\tilde{\sigma}^2}\right].
\end{align}
Notice that $\tilde{\sigma}$ is the packet width in the frequency space while $\omega_0$ is its central frequency. The initial condition gives us the desired Gaussian wave packet in the position space,
\begin{equation}
    |\psi(0,r^\star)|^2 = \frac{1}{\sqrt{2\pi}}\exp\left[{-\frac{r^{\star^2}}{2\sigma^2}+i\omega_0 r^\star}\right], \quad \sigma \equiv \tilde{\sigma}^{-1}
\end{equation}
but due to numerical limitations, we will be only interested in the interval $\omega\in(\omega_0-2\tilde{\sigma},\omega_0+2\tilde{\sigma})$ around the central frequency $\omega_0$, so that our integration is done as
\begin{equation}
    \psi(t,r^\star) = \int_{\omega_0-2\tilde{\sigma}}^{\omega_0+2\tilde{\sigma}} \tilde{\psi}(\omega)e^{-i\omega(t+r^\star)}\diff{\omega},
\end{equation}
with the same procedure applied to the transmitted and reflected waves. This was done so that the numerical integration was less time consuming, but the outcome is essentially the same since we are taking an interval of $4\tilde{\sigma}$ of a Gaussian packet. For what follows, we have chosen $\tilde{\sigma}=0.4\,\mbox{m}^{-1}$ (we recall that $\tilde{\sigma}$ is defined in the frequency space).

\begin{figure}[tb]
    \centering
    \includegraphics[width=0.95\linewidth]{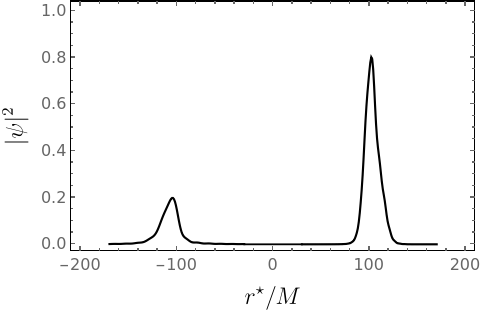}
    \caption{\label{fig:pkts}The transmitted (left) and reflected (right) wave packets in the position space for $\ell=5$, $Q=0.9M$ and $M\mu=1.0$. We can see that both packets have  pronounced peaks, located at $r_{tp}^\star$ for the transmitted packet, and $r_{rp}^\star$ for the reflected one. 
    This supports the integration over a finite $4\sigma$-interval around each peak and the use of the definitions for $T$ and $R$.}
\end{figure}

\begin{figure}[tb]
	\begin{flushleft}\textbf{(a)}\end{flushleft}
	
	\includegraphics[width=1\linewidth]{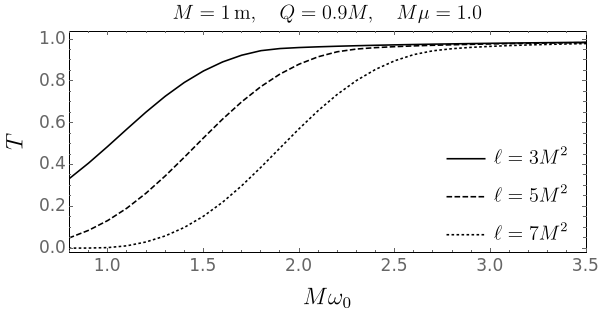}

	\begin{flushleft}\textbf{(b)}\end{flushleft}
	
	\includegraphics[width=1\linewidth]{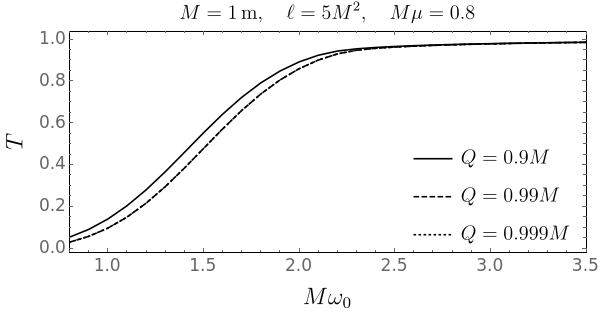}

	\begin{flushleft}\textbf{(c)}\end{flushleft}
	
	\includegraphics[width=1\linewidth]{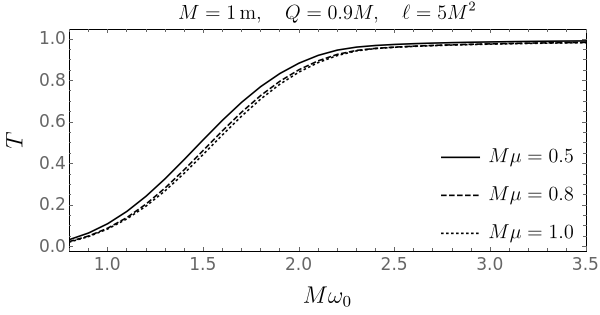}

	\caption{\label{fig:T}Plots for the transmission coefficient $T$ of the wave packet as a function of the central frequency $\omega_0$. In panel \textbf{(a)} we have the dependency of the $T$ for different values of the packet's angular parameter $\ell$; in panel \textbf{(b)} we have for different values of the BH's charge $Q$; while in panel \textbf{(c)} we have different values for the packet's mass $\mu$.}
\end{figure}

We then proceed on defining the reflection and transmission coefficients for the wave packet via
\begin{align}
    &R\equiv 
    \frac
    {\displaystyle\int_{r_{rp}^\star-2\sigma}^{r_{rp}^\star+2\sigma} |\psi_r(t_0,r^\star)|^2\,\diff{r^\star}}
    {\displaystyle\int_{r_{ip}^\star-2\sigma}^{r_{ip}^\star+2\sigma} |\psi_i(t_0,r^\star)|^2\,\diff{r^\star}},
    \label{eq:Rpkt}
    \\[1em]
    &T\equiv 
    \frac
    {\displaystyle\int_{r_{tp}^\star-2\sigma}^{r_{tp}^\star+2\sigma} |\psi_t(t_0,r^\star)|^2\,\diff{r^\star}}
    {\displaystyle\int_{r_{ip}^\star-2\sigma}^{r_{ip}^\star+2\sigma} |\psi_i(t_0,r^\star)|^2\,\diff{r^\star}}    
    \label{eq:Tpkt}
\end{align}
where $\psi_i(t,r^\star)$, $\psi_r(t,r^\star)$ and $\psi_t(t,r^\star)$ are the wave packets for the incoming, reflected and transmitted waves, respectively (note that $\psi_r$ and $\psi_t$ are not necessarily Gaussian, but it will be sufficient to have a pronounced peak at some asymptotic $r^*$ coordinate and concentrated around this peak in order for the definitions in eqs.~\eqref{eq:Rpkt} and~\eqref{eq:Tpkt} to be valid, as shown in fig.~\ref{fig:pkts}), and $t_0>0$ is an arbitrary fixed instant in time. The integration will also be done over an interval of $4\sigma$ around the peak of each packet (to be known, $r_{ip}^\star$, $r_{rp}^\star$ and $r_{tp}^\star$ for the incoming, reflected and transmitted packets, respectively). The results are shown in fig.~\ref{fig:T}.

In fig.~\ref{fig:T}, we can see from panel \textbf{(a)} that the transmission rate is sensible to the choice of the angular parameter $\ell$. This is expected since it is directly related to the impact parameter $b\sim (\ell+1/2)/M\omega_0$. In panel \textbf{(b)} we have different configurations of the charge $Q$ such that when $Q=0.99M$ and $Q=0.999M$ the lines become almost superposed, meaning that the difference for the transmission rates is almost negligible. Yet, they are still different as we are going to show in the next section. 

In all those configurations we have $M \gtrsim Q$ --- known as the \textit{near-extremal condition} for a RN BH. For panel \textbf{(c)}, we notice that different values of the mass $\mu$ of the packet act as a shifter for the transmission rate. In other words, when $\mu$ increases, the curve is displaced slightly to the right, meaning that less-massive (lighter) packets are more easily absorbed when compared to more-massive (heavier) packets with the same central frequency and parameters $M,Q,\ell$.

\section{Wave packet emission and CCC violation}
\begin{figure}[t]
    \centering
    \includegraphics[width=1\linewidth]{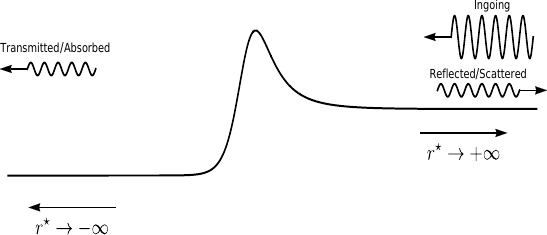}
    \caption{\label{fig:diag}Diagram of the incident, reflected and transmitted plane waves. The solid line represents the effective potential.}
\end{figure}

The absorption problem studied before can be seen diagrammatically in fig.~\ref{fig:diag}.
So far we have been studying the problem of an incident wave packet from $r^\star\to+\infty$ impinging on and being scattered by/transmitted to the BH. This means that in the given configuration the transmitted packet is actually absorbed by the BH (this comes from the mapping between $r$ and $r^\star$).

\begin{figure}[b]
    \centering
    \includegraphics[width=0.95\linewidth]{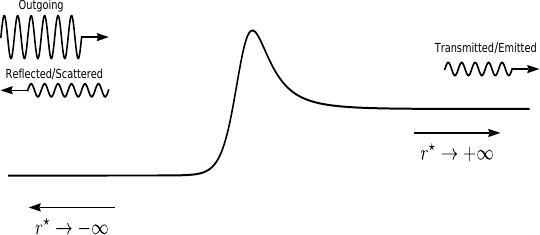}
    \caption{\label{fig:emission}Diagram of the outgoing, reflected and transmitted plane waves. Due to this inversion, the transmitted plane wave is now emitted by the BH.}
\end{figure}

Given the symmetry of the problem, we might as well think ``backwards'' in the sense that the absorbed packet becomes the emitted one, and the incident packet becomes the boundary conditions on measuring the emitted packet by the BH (see Fig.~\ref{fig:emission}). This is only possible due to a parity transformation of the transmission and reflection rates while interchanging $\omega$ with $\varpi$, regardless of the potential symmetry~\cite{Cohen-Tannoudji:101367}. This way, we may calculate the probability of a BH of mass $M$ and charge $Q$ to emit a neutral packet of mass $m_w$ such as the new mass $M'\equiv M-m_w$ violates the CCC with $M' < Q$ and gives us a naked singularity. Since the packets we have constructed here have mass around $m_w = 10^{-7}\,\mbox{eV}/c^2$, the difference between $M$ and $Q$ must be as small as possible. This also means that we are dealing with extremely light, axion-like particles~\cite{Ringwald:2014vqa}.

\begin{figure*}[t]
		\includegraphics[width=0.4\textwidth]{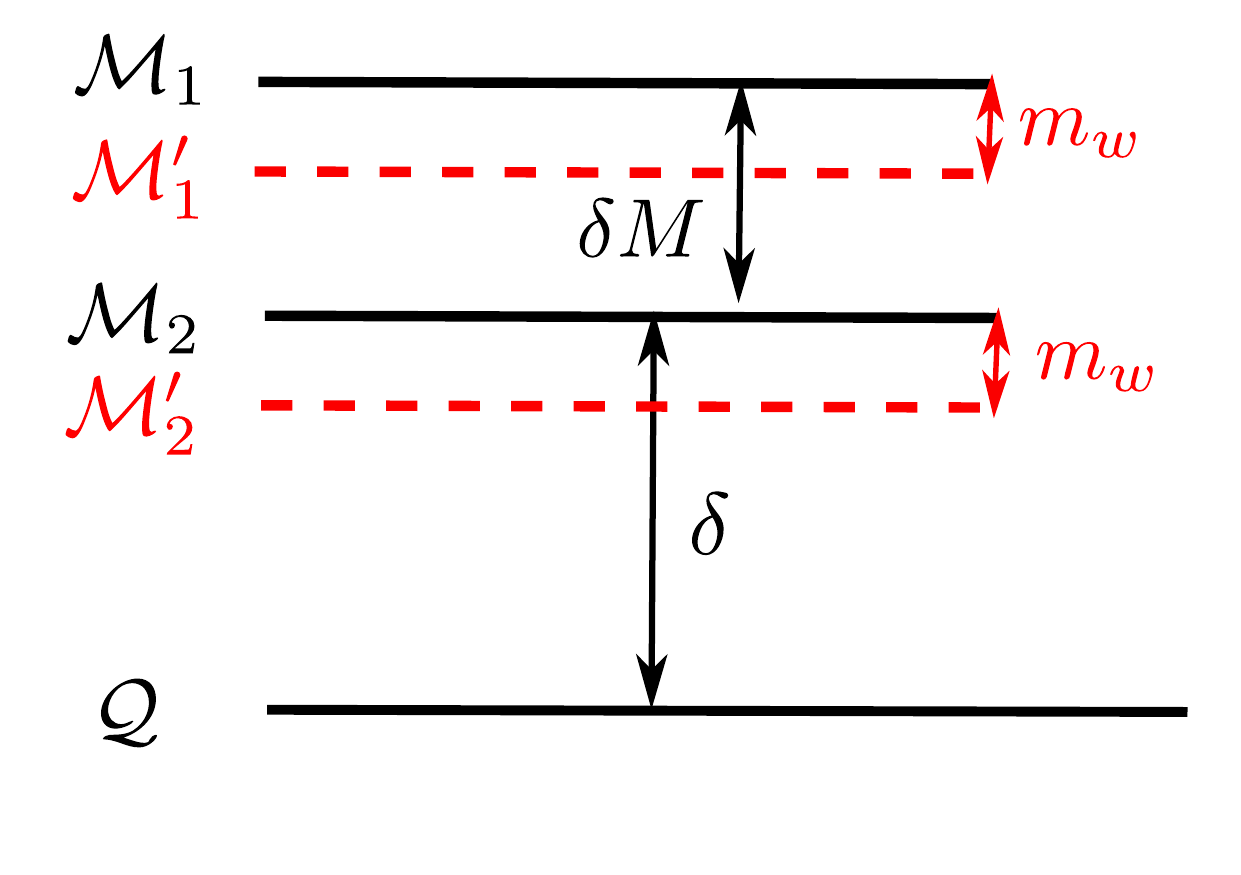}
		\includegraphics[width=0.4\textwidth]{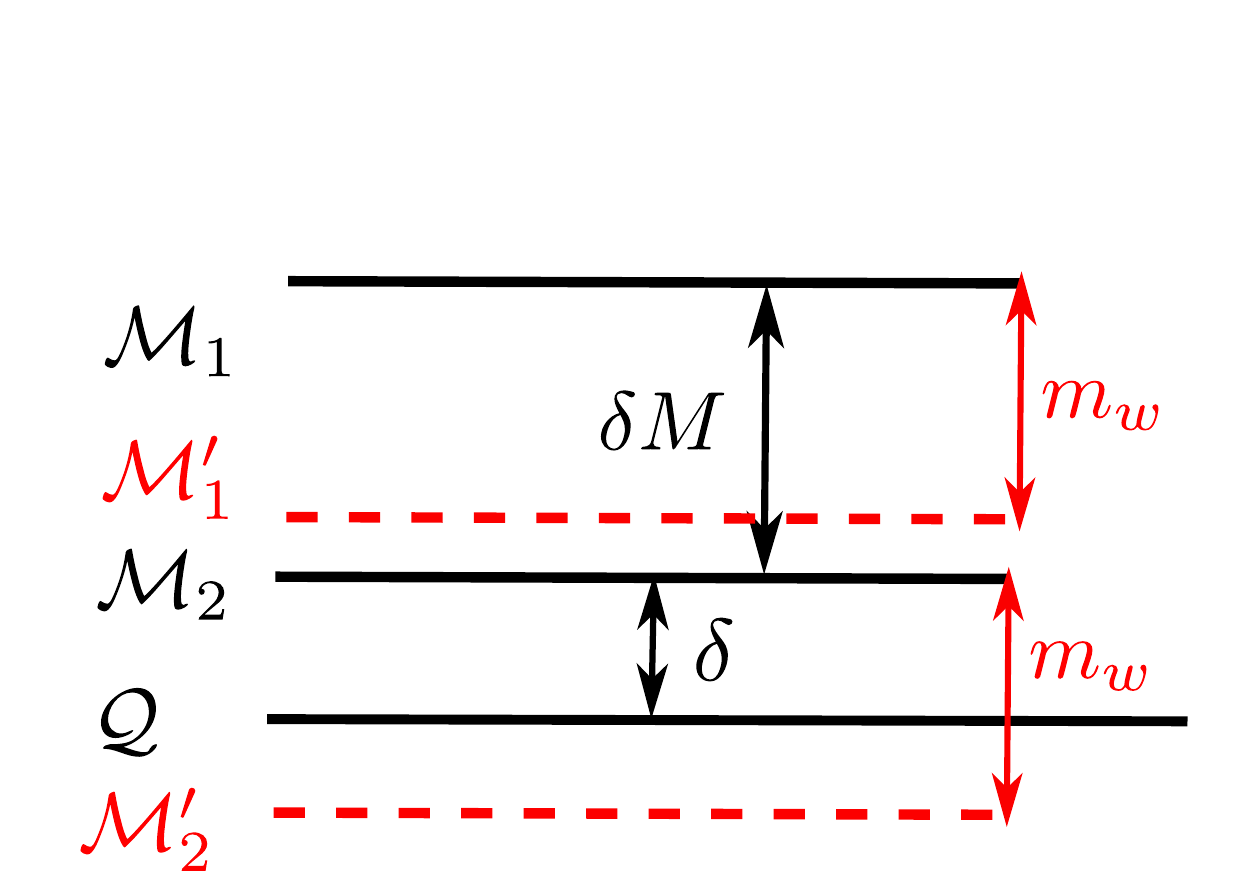}
		\caption{\label{fig:twolevel2}A ``two-level system'' scheme of the CCC for the two configurations of the BH. In the left panel we have $\delta > m_w$ meaning that the emission of a wave packet of mass $m_w$ imposes no danger to the CCC for both configurations. In the right panel we have $\delta \lesssim m_w$, meaning that for the second configuration, if $\mathcal{M}_2$ emits a wave packet of mass $m_w$ there will be a violation of the CCC. The difference between $\mathcal{M}_1$ and $\mathcal{M}_2$ in both images is $\delta M$.}
\end{figure*}

Let us consider two initial configurations for the BH. In the first one,  the mass will be $\mathcal{M}_1$ and charge $\mathcal{Q}$, while the second one the mass will be $\mathcal{M}_2$ and same charge $\mathcal{Q}$. Now, we want the difference between the masses of both configurations to be as small as possible, that is, $\mathcal{M}_1-\mathcal{M}_2 \equiv \delta M \ll M$, where $M$ is a base value for the mass of the BH, i.e, if we define $\mathcal{M}_1 \equiv M + \delta M$, then we will have $\mathcal{M}_2 \equiv M$. Next, we will impose that the diference between the mass $M$ and charge $\mathcal{Q}$ to also be as small as possible \footnote{We have also kept the same mass-charge ratio and the results are qualitatively the same.}, meaning that $M-\mathcal{Q} = \delta \ll M$, leading to $\mathcal{Q} \equiv M - \delta$. In this case, we have $\mathcal{M} \gtrsim \mathcal{Q}$, i.e., a near-extreme RN BH. We then define the ratio between their respective transmission rates as a function of $\mathcal{M}$ as
\begin{equation}\label{eq:TRatio}
    f_\epsilon \equiv \log\Bigg[\frac{T(\mathcal{M}_1)+\epsilon}{T(\mathcal{M}_2)+\epsilon}\Bigg].
\end{equation}
The cutoff factor $\epsilon$ is necessary to solve the convergence problem  when $T(\mathcal{M}_2)$ goes to zero faster than $T(\mathcal{M}_1)$ and the ratio diverges. Note that $\epsilon$ must be chosen as small as possible so it does not interfere with the values of $T$. Since $T$ ranges from $0$ to $1$, then $\epsilon \leqslant 0.1$ is a reasonable choice for $\epsilon$. We have verified that the graph of $f_\epsilon$ is robust to changes in the value of $\epsilon$ in the aforementioned range.

We use the parameter $f_\epsilon$ to analyze a possible violation in the CCC. If $f_\epsilon > 0$, it means that the former configuration $\{M+ \delta M,{\cal Q} \} $ is more likely to emit a particle when compared to the latter $\{M,{\cal Q}\} $. On the other hand, of course, $f_\epsilon < 0$ means the opposite: the latter configuration is more likely to emit a particle when compared to the former. If $f_\epsilon = 0$, the emission probabilities are the same regardless of $\delta M$. 

For tiny values of $m_w \sim \delta$, the lighter ${\cal M}_2$ BH can, in principle, emit a particle with mass $m_w$ and then become an overcharged BH. This idea is illustrated in fig.~\ref{fig:twolevel2}.

\begin{figure}[b]
    \centering
    \includegraphics[width=0.9\linewidth]{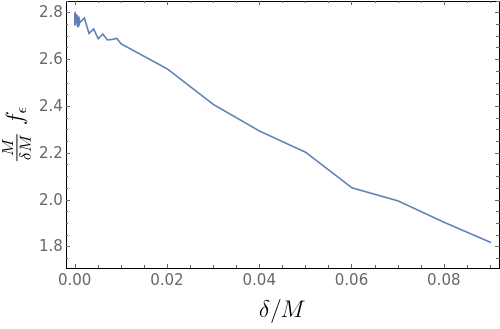}
    \caption{\label{fig:TRatio}Plot of $f_\epsilon$ as a function of $\delta$. Notice that as small as $\delta$ is chosen to be, the plot turns into noise. In this graph, we have $\delta M=10^{-11}M$. Also, $M\omega_0=1.0$, $\ell=5$ and $M\mu=0.8$.}
\end{figure}

Given the order of the mass of the packet, we must choose $\delta = 10^{-70}M$ (which is the order of the mass of the packet converted to geometrized unit system) to obtain a violation of the CCC, but we can see in fig.~\ref{fig:TRatio} that for very small $\delta$ (to be known, in the order of $\delta M$) will lead to use floating point precision for machine number and we will lose information. 

To sort out if for small $\delta$ the fixed value from which graph turns into noise has any physical significance, we have fitted the data within a reliable range of $\delta$ where our calculations ought to be correct. That said, in the range $\delta/M\in[10^{-3},\,10^{-1}]$ we found that $f_\epsilon$ may be written as
\begin{equation}\label{eq:fefit}
    f_\epsilon \approx f_\epsilon(\delta) = \frac{\delta M}{M}\Bigg(a \frac{\delta}{M} + b\Bigg),
\end{equation}
with $a=-10.7 \pm 0.4$ and $b=2.74 \pm 0.02$.

We point out that $\delta$ is the parameter which tells the closeness of the BH to become an extreme one. This result shows us that once both BHs are close to extremicity, the heavier BH is more likely to emit a particle of mass $m_w$ when compared to the lighter one.

Assuming that eq.~\eqref{eq:fefit} may be extended to smaller values of $\delta$, i.e, we may chose $\delta < m_w$ in order to achieve a violation of the CCC, according to the scheme in fig.~\ref{fig:twolevel2}.

We have seen above that there is a tiny, although finite, probability of tunneling and, therefore, overcharging a BH --- or so it seems. 
Nevertheless, that is not the only necessary ingredient to actually violate the CCC: one should also consider two essential factors. First, the typical discharging time, i.e, the time interval before a BH loses its charge (since it will attract opposite-charged particles from an orbiting cloud). Second, the travel time from the horizon to a typical radius (that may correspond to, say, the peak of the effective potential, eq.~\eqref{eq:Veff}, or to the radius circular orbit around the BH), which will provide --- very much like the old-fashioned description of a nuclear alpha decay~\cite{rohlf1994modern} --- the order of magnitude of the time interval between two successive ``attempts'' of such packet to escape.

We can estimate the former considering a massive BH such that the distance $\Delta r$ from $r=2M$ to $r=6M$ (or, as previously mentioned, any typical radius) is  classical, i.e, $\Delta r\gtrsim 10^{-10}\,\mbox{m}$. That requires $M\gtrsim 10^{-13}M_{\odot} \approx 10^{17}\,\mbox{kg}$ and, for a quasi-extremal case, $Q\gtrsim 10^7\,\mbox{C}$ --- which corresponds to about $10^{26}$ electrons or $10^{-5}\,\mbox{kg}$, much less than the BH mass. The proper time\footnote{The time-dilation factor that would yield the corresponding time for an observer at infinity is the same for both ingoing and outgoing particles, so it is not necessary to take it into account when comparing both.} of an outgoing packet (assuming it will be at rest at radial infinity) is about $10^{10}$ shorter than the ingoing time for the charged particle to cross the same distance (due to the attractive EM force). Therefore, a charged quasi-extremal BH would not remain so for the tunneling to take place.

We also recall that the ingoing charge does not need to tunnel, as opposed to the outgoing particle. Therefore, even if the discharging process takes, for some reason, the same amount of time of the outgoing flight, the tunneling probability of the outgoing packet --- which is $\sim 0.03$ for such massive BH  --- will decrease even further the rate of successful attempts.

\section{Discussion}
Among numerous papers and works, many of them have been done about waves being absorbed by the BH and how they could turn a near-extremal BH into a naked singularity, as seen in references~\cite{Glampedakis:2001cx,PhysRevD.89.104053,PhysRevD.52.1808,Andersson:2000tf,PhysRevD.18.1030,PhysRevD.91.124030,doi:10.1142/S0218271818430125}. Nevertheless, very little was done in the sense of a wave packet representing a particle being absorbed by the BH, alas being emitted by one. 

Even though this is a case study, in the sense that we are treating a neutral spin-$0$ particle. The method itself is really promising because all quantities here are analytical. The main idea of this work was to propose a feasible and reliable toy model to study the emission of particles by the BH without having to restrict ourselves to narrow frequency bands or particular numerical values. We have not only recovered all the results from previous researches, but we also have shown that this method gives a broader appliance, with all results being analytical and allowing broader (but not restricted) frequency bands. No approximations were needed to obtain any of the results shown here (apart from the obvious ones in the toy-model potential).

The asymmetry of the toy model is a key element to approach the problem discussed here once the actual effective potential is asymmetric itself. After we have constructed the toy model with two analytical functions to solve the Schr\"odinger-like eq.~\eqref{eq:SLE}, we were limited only to the choice of the set of parameters $\{M,Q,\ell,\mu\}$ and their physical limits. Of course, we could deal with a realistic case of a charged massive packet, but that implies that the frequency of the field would couple to its own charge in the effective potential, which would no longer be independent of the field's frequency. Therefore, the very construction of the wave packets themselves would become troublesome. Nonetheless, the method itself is a good approximation to the actual potential and many qualitative and quantitative information can be extracted from here.

Dealing with a localized wave packet may be a way of solving the problem of \textit{backreaction}: the initial conditions of the problem change. Nevertheless, the plane-wave approach corresponds to an eternal and constant flux of particles --- that is in direct opposition to the qualitative change in the metric (the violation of the CCC) which is being searched for.
That leads to numerous critics about given solutions to the violation of the CCC --- namely, that the effect of backreaction must be fully taken into account in order to perceive the problem in its entirety.

The semiclassical approximation of a particle --- namely, a wave packet --- can tell the probability of absorption each time it is thrown at a BH, since it is localized in the position space. We also recall that we have worked in the ``safe side'' of the metric, i.e., in the presence of an EH and outside it. We have shown that, within the numerical precision achieved, the CCC holds even when we start from closer and closer to an extreme configuration

It is important to point out that we did not consider the superradiance effect, where $R>1$. We recall~\cite{PhysRevD.93.024028} that the superradiance conditions, for a wave with frequency $\omega$ and charge $q$, are $q Q>0$ and $\omega < q Q/r_+$, with $q$ being the particle's charge, In this paper we have $q=0$, thus the conditions become null and we have no superradiance.

Eq.~\eqref{eq:TRatio} is the main driver to our conclusions regarding CCC violation. As it was mentioned in the previous section, $f_\epsilon$ tells us the probability ratio of a particle to be emitted between two given initial configurations. The graph in fig.~\ref{fig:TRatio} showed us that for small values of $\delta$ a more massive BH is more likely to emit a particle when compared to a lighter one. For very small values of $\delta$ the graph of $f_\epsilon$ becomes noisy due to floating-point precision, and for that we have fitted a model within a reliable range of $\delta$ close to zero in eq.~\eqref{eq:fefit} which showed that $f_\epsilon > 0$, meaning that this tendency is kept throughout the process.

Even though there is a tiny --- but non-null --- chance for the particle to be emitted and violate the CCC by exposing the singularity, we have calculated the time required for such emission to happen and compared it to the typical time for the BH to capture an external charge and increase the $\delta$ difference. We have found out that the former is greater than the latter. This means that, even though a particle may be emitted and violate the CCC, the BH will capture a charged particle faster than the emission process occurs, which increases the $\delta$ difference and leaves the BH further from extremicity for the time when the emission takes place.

\section{Conclusions}
Our method shows promising results regarding the particle emission from a BH via quantum tunneling process. The toy model ended up as an extremely useful tool to avoid frequency approximations and numerical analysis as the only means to obtain the desired answers for a large range of the parameters of the system. The simplicity of the method allows the use of an everyday computer with no particular configuration to calculate complex quantities in matters of minutes with good approximation and achieve frequency bands as broader as the machine can reach. 

Using wave packets to represent particles is a novelty in the sense that, up to today, no work has applied this approach in the violation of the CCC. Besides, only a few papers have applied toy models to BHs, but in the sense of its properties and stability in the near-extremal case~\cite{Maldacena:2017axo,Maldacena:2016upp,Maldacena:2016hyu}.

In general, as a case study, the method itself and its results are indeed promising, where much more can be done from it. The next step is to apply the same procedure to a neutral, spinning BH.

\section{Acknowledgements}
R. L. F. would like to thank CAPES for the financial support under the grant process number 88882.331077/2019-01 throughout this work. S. E. J. thanks Fernando D. Sasse for fruitful discussions on the subject.

%
\end{document}